\documentclass[journal]{IEEEtran}
\ifCLASSINFOpdf
\else
\fi

\usepackage{graphicx}
\usepackage{amsmath}
\usepackage{datetime2}
\usepackage{hyperref}
\usepackage{caption}
\usepackage{subcaption}

\usepackage{breakurl}

\begin{document}
%
\title{Epidemic modelling of multiple virus strains: \\a case study of SARS-CoV-2 B.1.1.7 in Moscow}
%
%
%


\author{Boris~Tseytlin,
        Ilya~Makarov\\
        b.tseytlin@lambda-it.ru, iamakarov@hse.ru\\
        HSE Univercity, Moscow
}
\maketitle

\begin{abstract}
During a long-running pandemic a pathogen can mutate, producing new strains with different epidemiological parameters. Existing approaches to epidemic modelling only consider one virus strain. We have developed a modified SEIR model to simulate multiple virus strains within the same population. As a case study, we investigate the potential effects of SARS-CoV-2 strain B.1.1.7 on the city of Moscow. Our analysis indicates a high risk of a new wave of infections in September-October 2021 with up to 35 000 daily infections at peak. We open-source our code and  data\footnote{\url{https://github.com/btseytlin/covid_peak_sir_modelling}}.
\end{abstract}


%
\IEEEpeerreviewmaketitle

\section{Introduction}

\IEEEPARstart{I}{n} March 2020, the World Health Organization (WHO) declared COVID-19, a disease caused by the SARS-CoV-2 virus, a global pandemic \cite{whostatement}. Up until August 2020 the number of worldwide COVID-19 cases doubled approximately every 20 days. As of April 2021, there have been over 130 million confirmed cases and over 2.8 million deaths. 

Since the start of the epidemic SARS-CoV-2 has mutated into multiple new major strains. Some of these strains have the potential to change the course of pandemic and were deemed variants of concern (VOC) by the World Health Organization. As of time of writing, the list of VOC includes the B.1.1.7 strain from United Kingdom, the B.1.351 from South Africa and the B.1.1.28.1 \cite{whositrep}. The B.1.1.7 strain has been reported to have a 40\% - 90\% larger basic reproduction number than the base SARS-CoV-2 \cite{davies2021estimated,volz2021transmission} and it has spread to 130 countries at the time of writing \cite{whositrep}. In combination this makes the B.1.1.7 potentially dangerous and it is important to assess the risks and plan the response.


Epidemic modelling was successfully used to estimate the spread of the virus and provided tools to plan the optimal response strategy \cite{roda2020difficult}. In particular, modifications of the SEIR model \cite{kermack1927contribution} were effective, as both simple and capable of simulating a complex epidemic \cite{calafiore2003modified,ahmad2020report,sun2020forecasting,sameni2020mathematical,thompson2020time,kucharski2020early,asatryan2020predicting}. There were multiple attempts to apply more advanced machine learning techniques to COVID-19 forecasting \cite{zheng2020predicting,dandekar2020quantifying,tian2020forecasting}. However, Sun et al. report that due to insufficient training data and over-fitting more advanced machine learning models have not achieved improvements significant enough to justify their added complexity \cite{sun2020forecasting}. 

The core research question of our work is not to provide the most accurate forecast, but to estimate the potential impact of a new virus strain. We chose a SEIR-based approach to this problem for the following reasons. First, the parameters of SEIR models are simply characteristics of the disease, which allows to use medical studies to constrain parameter ranges. Second, Unlike black-box models, SEIR models provide the ability to adjust parameters and simulate multiple "what-if" scenarios. This is crucial for simulating what a virus strain with a 40\% - 90\% larger basic reproduction number can do.



The main contributions of our work are as follows:
\begin{enumerate}
    \item We develop a modified SEIR model that accurately simulates the SARS-CoV-2 epidemic in the city of Moscow while taking in account under-reporting in data and quarantine measures.
    \item We propose an an easy yet effective function to model quarantine measures.
    \item We extend the model for multiple virus strains and forecast scenarios of B.1.1.7 spread in Moscow.
\end{enumerate}

The remaining paper is structured as follows. Section \ref{sec:seird_background} provides necessary background on SEIR models. In Section \ref{sec:covid_model} we describe the proposed SEIR modification and modelling quarantine measures. Section \ref{sec:experiments} describes the evaluation process, dataset, parameter optimization and the resulting accuracy metrics. Finally, in Section \ref{sec:two_strain} we extend the model for multiple strains. In Section \ref{sec:forecast_moscow} we obtain forecasts for B.1.1.7 in Moscow.

\section{SEIR background} \label{sec:seird_background}
The original SEIR model \cite{kermack1927contribution} simulates an epidemic in a closed population of size $N$. The SEIRD model \cite{bailey1975mathematical} extends SEIR by adding fatalities from infection. We use the SEIRD model further on, as it is more relevant to our task.

In SEIRD, the population is partitioned into five compartments: Susceptible, Exposed, Infections, Recovered and Deceased. Susceptible individuals can be infected. Exposed individuals have been infected, but are not spreading the pathogen yet. Infectious individuals are spreading the pathogen. Recovered individuals are permanently immune.  The size of compartments at time $t$ is denoted by $S(t)$, $E(t)$, $I(t)$, $R(t)$, $D(t)$. The following ODE system describes the transfer of population between compartments:
\begin{equation} \label{seird}
    \begin{split}
        \frac{dS}{dt} & = -\frac{\beta S(t) I(t)}{N} \\
        \frac{dE}{dt} & = \frac{\beta S(t) I(t) }{N} - \delta E(t) \\
        \frac{dI}{dt} & = \delta E(t) - \gamma (1 - \alpha) I(t) - \gamma \alpha I(t) \\
        \frac{dR}{dt} & = \gamma (1- \alpha) I(t) \\
        \frac{dD}{dt} & = \gamma \alpha I(t) 
    \end{split}
\end{equation}

subject to initial conditions $S_0, E_0, I_0, R_0, D_0$ and the constraint $N = S(t) + E(t) + I(t) + R(t) + D(t)$. The parameters of the model are: $\alpha$ - infection fatality rate, $\beta$ - number of cases generated by an infected individual in one day, $\delta = 1/d_{incubation}$ where $d_{incubation}$ is the length of incubation period, $\gamma = 1/d_{infectious}$ where $d_{infectious}$ is the time in days till recovery or death.  

The basic reproduction number $R_0$ is the defining characteristic of a pandemic. It equals the expected number of people infected by one infectious individual during the course of their sickness and can be computed as follows:
\begin{equation} \label{r0}
    R_0 = \beta / \gamma
\end{equation}

\section{Modelling the COVID-19 epidemic in Moscow} \label{sec:covid_model}

To create an epidemic model for two strains we first have to create an accurate model of the one-strain epidemic. SEIRD requires modifications to simulate a long-running pandemic like COVID-19. It assumes all parameters to stay constant, but this assumption is violated by quarantine measures, where $R_0$ changes with time. Optimizing the parameters on historical data is also problematic as the model ignores possible under-reporting. To resolve these issues we modify the SEIRD model with a quarantine measures function and additional compartments.

\subsection{Modelling quarantine measures}
\label{sec:quarantine}
In case of COVID-19 quarantine and social distancing measures directly affect the number of contacts and the probability of infection on contact, which makes $\beta$ dependant on measures currently in effect. As a consequence, a model with constant $\beta$ is not able to simulate multiple waves of epidemic. To model the effect of quarantine measures we replace the constant parameter $\beta$ with a time-dependant function $\beta(t)$. 

For $\beta(t)$ we propose a simple model where at every $t$ the basic reproduction number is reduced by a factor of $q(t) \in [0, 1]$.
Then we can then obtain $\beta(t)$ from $R(t)$ using $\autoref{r0}$.
\begin{equation} \label{r(t)}
    R(t) = R_0 - R_0 \cdot q(t)
\end{equation}
\begin{equation} \label{beta(t)}
    \beta(t) = R(t) \cdot \gamma
\end{equation}

In general $q(t)$ can be any differential function. We used a stepwise function with sigmoid transitions between steps: 
\begin{equation} \label{q(t)}
q(t) = \begin{cases} 
      q_1 & t \leq t_{1} \\
      transition(t, t_1, t_2, q_1, q_2) & t_{1} < t \leq t_{2} \\
      \dots \\
      transition(t, t_{i-1}, t_{i}, q_{i-1}, q_{i}) & t_{i-1} < t \leq t_{i} \\
      \dots \\
      q_n & t \geq t_{n}
   \end{cases}
\end{equation}

\begin{equation} \label{sigmoid}
\begin{split}
& transition(t, t_{a}, t_{b}, q_a, q_b) = \frac{q_a \exp(s r) + q_b  \exp(r t_s)}{ \exp(s r) + \exp(r t_s ) } \\
& t_{s} = (t - t_{a}) / (t_{b} - t_{a})
\end{split}
\end{equation}

The $q_1, q_2, \dots, q_n$ values are quarantine power values to be obtained by optimization and the knot points $t_1, t_2, \dots, t_n$ are hyperparameters. The $transition$ function produces a soft descent or ascent from $q_a$ to $q_b$. The parameters $s$ and $r$ control the shape of transition. Empirically we found $r=20$ and $c=0.5$ to work well for our case.

\subsection{SEIRD-H: modelling under-reporting in data}

In reality statistical data on infections, deaths and recoveries is not complete. Many people go through the sickness without getting tested or hospitalized. Errors are possible when registering deaths as well. For example, a previous study \cite{kobak2021excess} has shown that in Moscow the real death toll from COVID-19 might be 2.6 times higher than what the official statistics shows. We propose a modification to SEIRD to take in account the incompleteness of statistical data. 

Let cases of infection, death and recovery be called \textit{visible} if they are registered in statistical data, and \textit{invisible} if they are undetected. Upon infection a case becomes \textit{visible} with probability $p_i$, and \textit{invisible} with probability $1 - p_i$. A death of an \textit{invisible} case is registered as a \textit{visible} death with probability $p_d$ to handle the situation where a person dies from unknown causes and is then revealed to have died from COVID-19.

\begin{figure}[t]
    \centering
    \includegraphics[width=\linewidth]{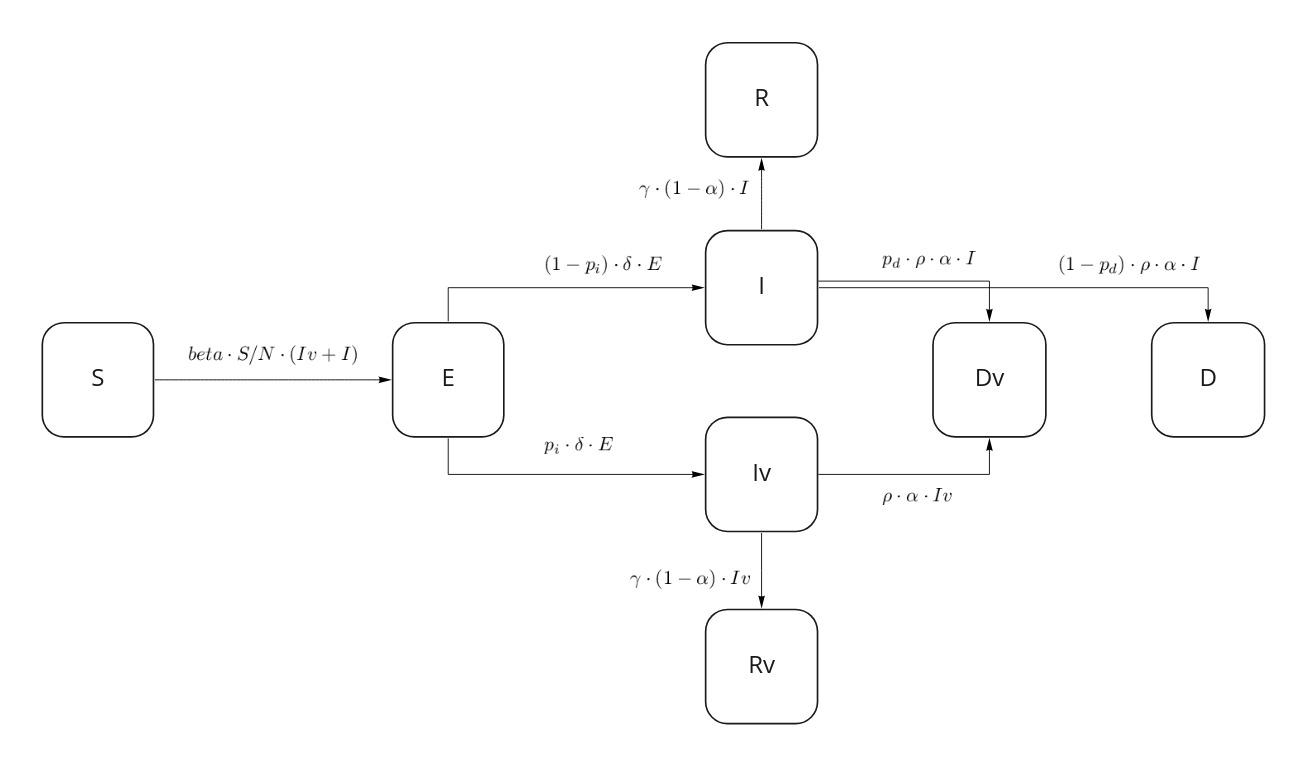}
    \caption{SEIRD-H: SEIRD with hidden states.}
    \label{fig:seird_hidden}
\end{figure}

To model these relationships we modified the compartments of the SEIRD model. In the modified model compartments $I_v$, $R_v$, $D_v$ contain \textit{visible} cases and compartments $I$, $R$, $D$ contain \textit{invisible} cases. We nickname this model SEIRD-H: SEIRD with hidden states. The ODE system for the SEIRD-H model is provided in Appendix \ref{appendix:ode_systems} \autoref{eq:seridh_one_strain} and \autoref{fig:seird_hidden} shows the schematic of the new model.

How do we obtain the initial conditions $S_0$, $E_0$, $I_0$, $I_{v_0}$, $R_0$, $R_{v_0}$, $D_0$, $D_{v_0}$? The sizes of invisible compartments can not be directly observed. Our solution is to run the simulation from the first infected case till the required date. Consider the case of Moscow data: the first date of dataset is 2020-03-12, while the first official recorded COVID-19 case in Moscow is on 2020-03-02. We set the compartment values to $S=N-1, I=1$, and the rest to zeros, and run the simulation for ten days. The obtained states are the initial conditions for further simulation.

We have established a modified SEIR model by adding a time-dependant function of quarantine measures and hidden compartments for cases invisible in statistical data. However the model is not complete until we find such parameters that it's able to accurately simulate the one-strain epidemic. Additionally, the accuracy of the model must be experimentally tested before it can be extended for virus two-strains. 

\section{Experiments}
\label{sec:experiments}
We aim to produce a model that outputs values for visible $I$, $D$ and $R$ cases as close as possible to statistical data. However it's not enough to check the fit of our model on training data. In reality we want to train the model on historical data and produce forecasts for the feature. We have to use an evaluation procedure that imitates this regime to get a robust estimate of the quality of forecasts. One way to do this is time-aware cross-validation. In this setting, an evaluation date s randomly selected. The model parameters are optimized on historical data before this date. The model produces a forecast for the time range after the evaluation date. The forecast is compared to the ground truth data and error is computed. The process is repeated many times and final error metrics are averaged.

We evaluate our model using time-aware cross validation and use the Mean Absolute Error (MAE) and Symmetric Mean Percentage Error (SMAPE) metrics. The results are compared to two baselines: the persistence model, which always predicts the last training value, and the unmodified SEIRD model.

\subsection{Dataset}

For conducting the experiments the official COVID-19 data provided by Moscow authorities and made available by Yandex \cite{yandexstat} was used. The dataset spans the time range between \DTMdisplaydate{2020}{3}{10}{-1} and \DTMdisplaydate{2021}{3}{23}{-1}. For each date the following statistics were provided: new confirmed cases, new recoveries, new deaths, total confirmed cases, total recoveries, total deaths.

\subsection{Optimizing model parameters}

We optimized the model-predicted \textit{new daily visible} infections, deaths and recoveries to be as close as possible to the statistical data. The Levenberg-Marquardt method via the lmfit \cite{newville2016lmfit} library was used for performing the optimization. Ranges of possible parameter values were constrained using information from COVID-19 studies \cite{10.3389/fpubh.2020.598547,asatryan2020predicting}.

\begin{table}[t]
\centering
\caption{Parameters of the one-strain model obtained by optimization}
\label{table:parameters}
\begin{tabular}{ |c|c|c|c|c|c|c|c|c|c| } 
 \hline
Parameter & Range of values &  Obtained value\\ 
\hline
$R_0$ & [3, 5] &  4.781 \\ 
$\alpha$ & [5e-3, 7.8e-3] &  6.4e-3  \\ 
$\delta$ & [1/14, 1/2] &  1/2 \\ 
$\gamma$ & [1/14, 1/7] &  1/9 \\ 
$p_i$ & [0.15, 0.3] &  0.25 \\ 
$p_d$ & [0.15, 0.9] &  0.35 \\ 
$q_{60}$ & [0, 1] &  0.69 \\ 
$q_{120}$ & [0, 1] &  0.869 \\ 
$q_{180}$ & [0, 1] &  0.715 \\ 
$q_{240}$ & [0, 1] &  0.712 \\ 
$q_{300}$ & [0, 1] &  0.766 \\ 
$q_{360}$ & [0, 1] &  0.760 \\ 
\hline
\end{tabular}
\end{table}

 The final loss function is a sum of residuals for infections, deaths and recoveries. To put all residuals on the same scale we transformed them to symmetric relative error:
\begin{equation} \label{residuals}
    rel(true, pred) =\frac{true - pred}{|true| + |pred|}
\end{equation}


The final loss function $L$ is a weighted sum of relative errors for daily infected, recovered and dead:
\begin{equation} \label{loss}
    \begin{split}
    L_I = \sum_{t=1}^{n} rel(true_I(t), pred_I(t))\\ 
    L_R = \sum_{t=1}^{n} rel(true_R(t), pred_R(t))\\ 
    L_D = \sum_{t=1}^{n} rel(true_D(t), pred_D(t))\\
    L = w_I L_I + w_R L_R + w_D L_D
    \end{split}
\end{equation}
Where $n$ is the length of training time range, $true_I(t), true_R(t), true_D(t)$ are ground truth daily infections, recoveries and deaths on day $t$, and $pred_I(t), pred_R(t), pred_D(t)$ are output by the model, $w_I$, $w_R$, $w_D$ are hyperparameter weights for loss components. 

Empirically we found weights $w_D = 0.5, w_I = w_R = 0.25$ to work well.

\begin{figure}[t]
    \centering
    \includegraphics[width=\linewidth]{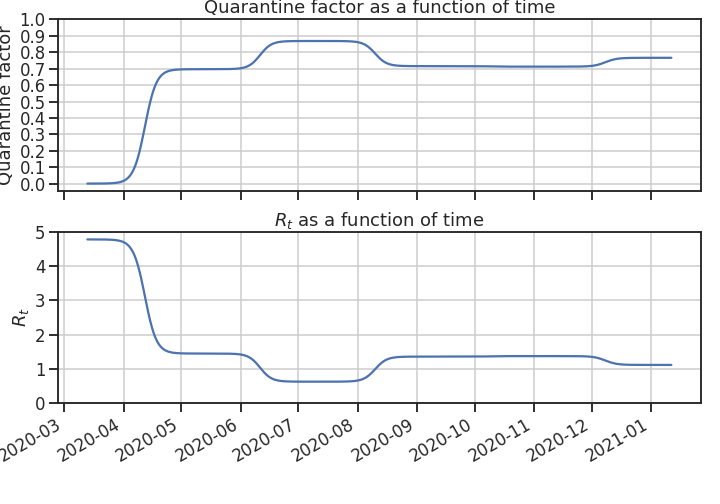}
    \caption{The effect of quarantine measures on $R_0$ as learned by the one-strain SEIRD-H. Top: quarantine power as a function of time. Bottom: basic reproduction number after quarantine effects as a function of time.}
    \label{fig:quarantine_function}
\end{figure}

\begin{figure}[t]
    \centering
    \includegraphics[width=\linewidth]{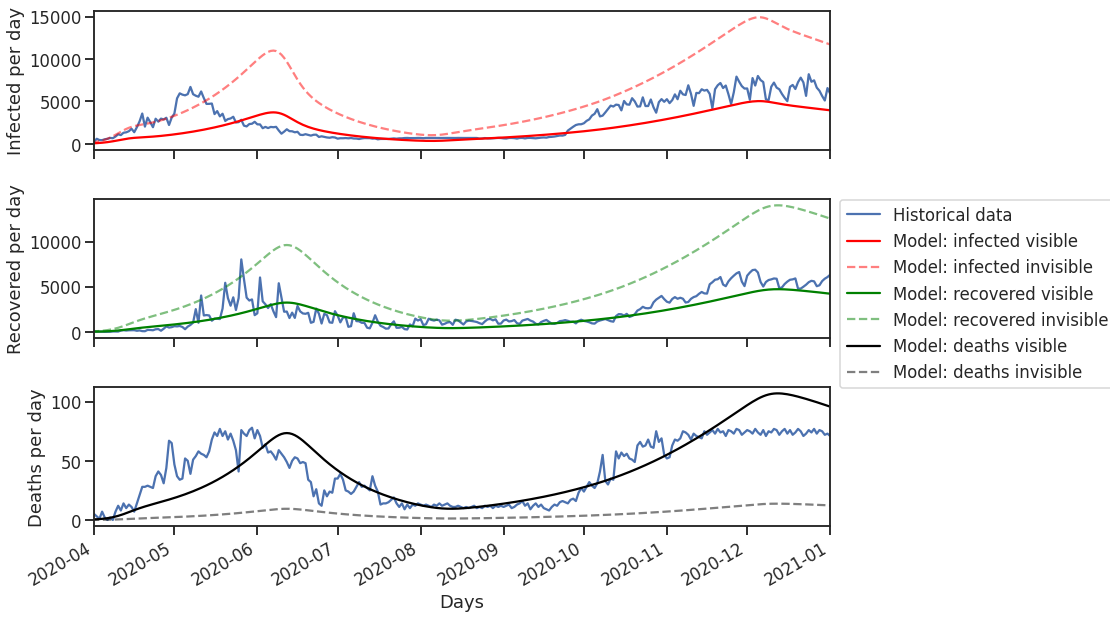}
    \caption{Output of one-strain SEIRD-H versus the ground truth on training data.}
    \label{fig:daily_infected_dead_recovered_train}
\end{figure}

The obtained parameters are provided in \autoref{table:parameters}. An interesting observation is that our models considers 75\% infectious cases to go undetected.  

The resulting quarantine measure function and training fit are presented on \autoref{fig:quarantine_function} and  \autoref{fig:daily_infected_dead_recovered_train}. The model predicts no quarantine measures at the beginning of the epidemic, a summer period of lockdown and a following period of eased quarantine measures, which approximately matches the real timeline in Moscow.

\subsection{Model verification}
\label{sec:verification}
For testing the one-strain model by time-aware cross-validation we pick 14 evenly spaced dates in the training range. For each date we train the model on the previous data, forecast cumulative deaths for 30 days ahead, and compute the MAE and SMAPE. 

\begin{table}[t]
\centering
\caption{Experiment results: cumulative deaths prediction errors.}
\label{table:metrics}
\begin{tabular}{ |c|c|c| } 
 \hline
Model &  MAE & SMAPE \\ 
\hline
Persistence & 715 & 5.00\% \\ 
SEIRD & 17072 & 36.7\% \\ 
SEIRD-H & \textbf{550} &  \textbf{4.20\%} \\ 
\hline
\end{tabular}
\end{table}

The resulting metrics are presented in \autoref{table:metrics}. Basic SEIRD fails to outperform the persistence baseline, mostly due to producing very bad predictions during the second wave of infections, as it's not able to simulate multiple waves. SEIRD-H outperforms both baselines. We conclude that SEIRD-H produces accurate enough predictions to extend it for modelling two virus strains. 

\section{Modelling two virus strains}
\label{sec:two_strain}

\begin{figure}[t]
    \centering
    \includegraphics[width=\linewidth]{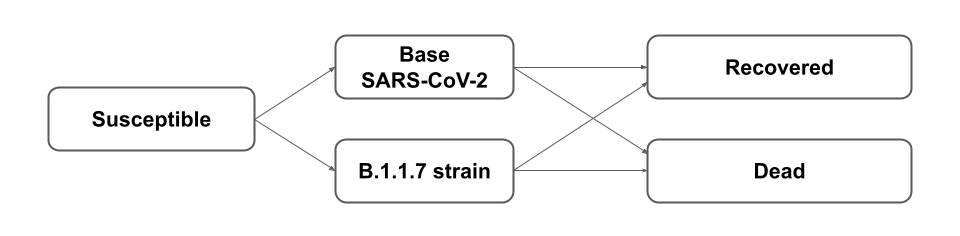}
    \caption{Two-strain model as two parallel pandemics that share the compartments of Susceptible, Recovered and Dead.}
    \label{fig:two_strain_highlevel}
\end{figure}

We can think of two strains within the same population as two parallel pandemics, with the condition that a person can only get infected once. This can be modelled by two SEIRD models that share the compartments of Susceptible, Recovered and Dead individuals. \autoref{fig:two_strain_highlevel} illustrates this idea.

The base SARS-CoV-2 model can be obtained by training a SEIRD-H model. For training this model we use historical data prior to the date of first infection by the second strain.

Obtaining the model for the second strain is trickier, as we do not have per-strain statistical data to train a separate model. We can use the fact that a new strain at it is core is the same disease as the base pathogen, but with slightly different epidemiological parameters. For example, we can assume that quarantine measures affect the new strain in the same way as the base strain. Thus, we can reuse the learned parameters of the base model to create the new strain model. 

The major difference between the base SARS-CoV-2 and B.1.1.7 is that B.1.1.7 has a 40\% - 90\% larger basic reproduction number \cite{davies2021estimated}. Then the B.1.1.7 model can be obtained by copying the parameters from the base model and modifying $R_{B.1.1.7}(t)$. For example, if B.1.1.7 is 50\% more infectious than the base strain then we can set $R_{B.1.1.7}(t) = 1.5 \cdot R(t)$.

The ODE system for the two-strain model is provided in Appendix \autoref{appendix:ode_systems} \autoref{eq:seridh_two_strains} and the detailed schematic is provided in Appendix \autoref{appendix:seird_two_strain} \autoref{fig:seird_two_strains}.

Now that we have obtained a model for two virus strains, we can use it to forecast the potential effect of B.1.1.7 in Moscow.

\section{Forecasts for B.1.1.7 spread in Moscow}
\label{sec:forecast_moscow}

\begin{figure}[t]
    \centering
        \includegraphics[width=\linewidth]{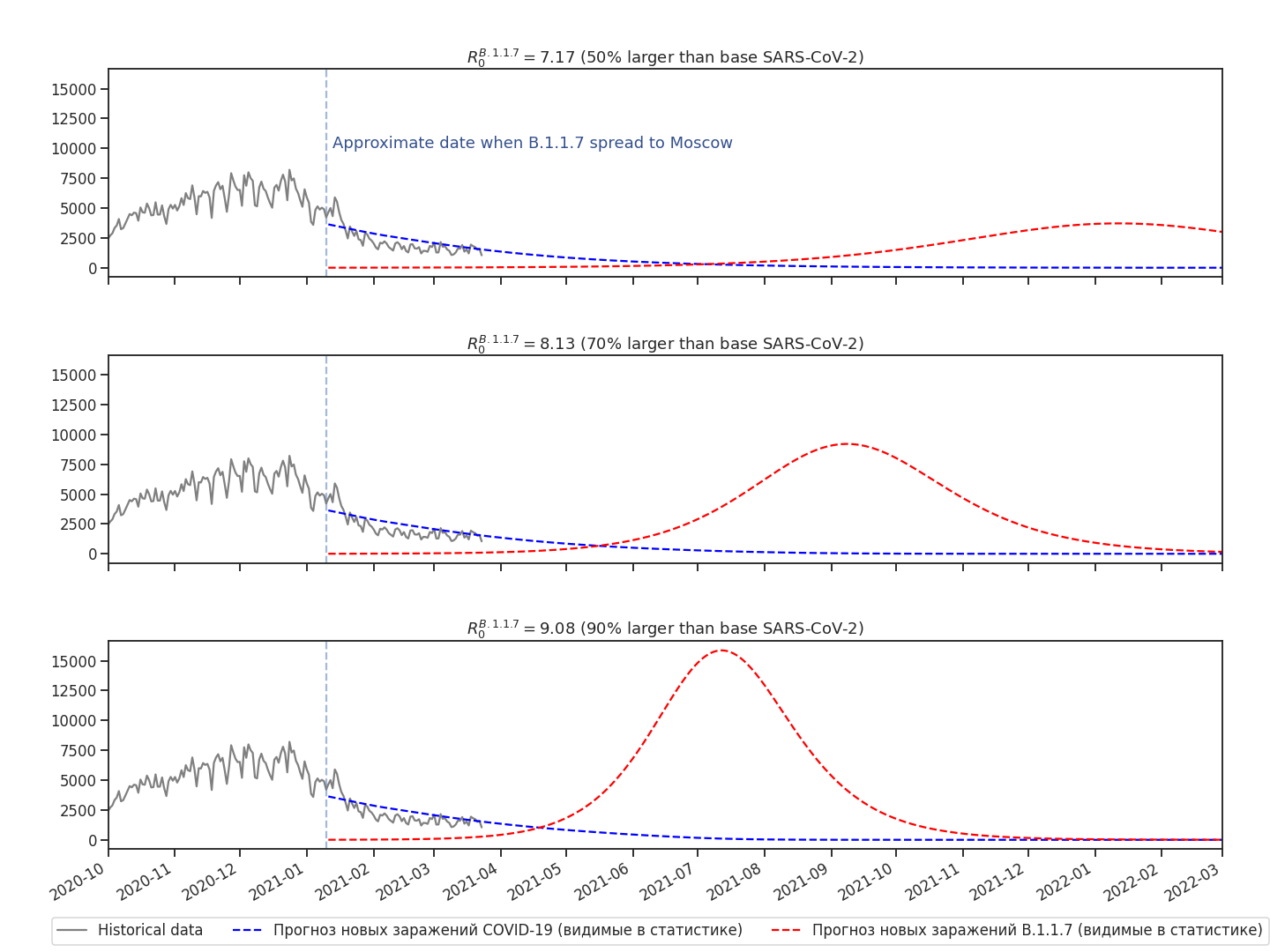}
        \caption{Forecasted scenarios depending on the $R_0$ of B.1.1.7. In the worst case of $R_0^{B.1.17} = 9.08$ the new strain causes a large new wave of infections. In the best case the new wave is still present, but is approximately of the same magnitude as the December 2020 infection levels.}
        \label{fig:scenarios_beta2mult}
\end{figure}

\begin{figure}[t]
    \centering
        \centering
        \includegraphics[width=\linewidth]{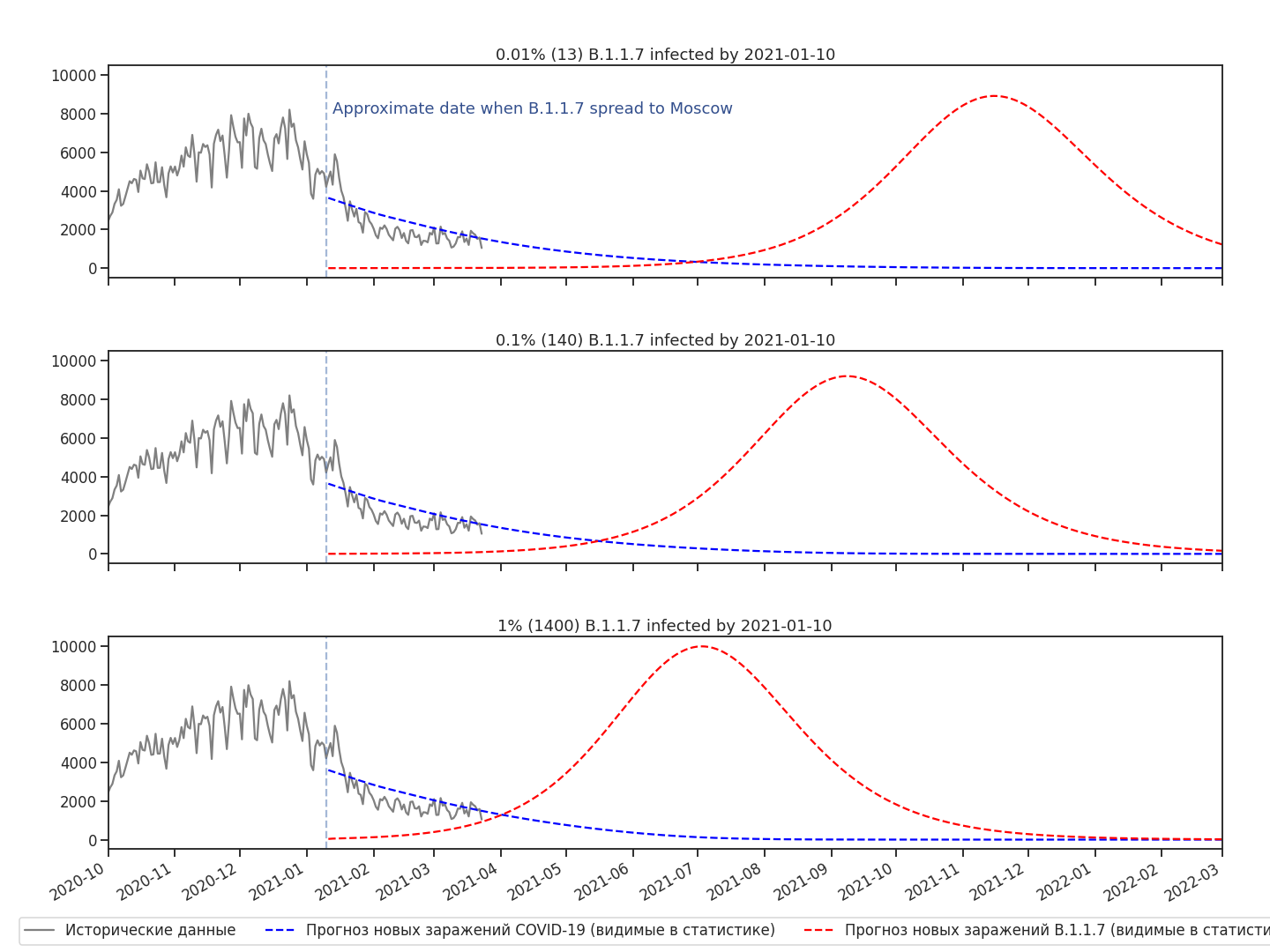}
        \caption{Forecasted scenarios depending on the ratio of new strain. The number of initial infectious B.1.1.7 cases only changes the peak date of new wave.}
        \label{fig:scenarios_new_strain_ratio}
\end{figure}

To obtain a forecast using the two-strain SEIRD-H we need to deal with two unknown factors: the true $R_{B.1.1.7}$ and the initial number of infectious B.1.1.7 cases. We do not know the exact values, so we have to investigate multiple scenarios.

On \autoref{fig:scenarios_beta2mult} we consider the scenarios for different $R_0$ values of B.1.1.7. For these scenarios we assume that by 2021.01.10 0.1\% of infectious are carrying the B.1.1.7 strain. The simulation indicates that a new wave is likely to happen for any possible $R_0^{B.1.1.7}$. 

\autoref{fig:scenarios_new_strain_ratio} represents scenarios for various numbers of B.1.1.7 strain carriers at 2021.01.10. For these scenarios it is assumed that $R_0^{B.1.1.7} = 8.13$(70\% greater than the base $R_0$). In the best case, if there are only 0.01\% (13) new case carriers, the new wave peaks in November-December 2021. In case of 1\% (1400) new case carriers, the new wave peaks in July 2021.

We consider the most likely scenario to be when $R_0^{B.1.1.7} = 8.13$ and the initial number of new strain carriers is 0.1\% (140). In that case the forecasted new wave starts developing in June 2021 and peaks in September-October 2021. At the peak it causes up to 35 000 new daily infections, out of which only 10 000 are detected in statistics. 

To sum up our analysis, the new wave of infections happens in every scenario, even though the magnitude and date of the peak vary. The new wave might be averted by a factor not considered in the model, like vaccinations or seasonality, but we can say for certain that B.1.1.7 possesses a significant risk. Up until July 2021 it will look like COVID-19 infections are declining, as the few B.1.1.7 will be invisible among numerous base strain cases. It will be tempting to relax quarantine measures. Then B.1.1.7 cases can explode exponentially and catch society by surprise. The bare minimum recommendation is to prepare for the explosion of cases in June-August 2021 and study B.1.1.7 further.

\section{Conclusions}

The aim of our work was to develop a model to asses the risk of a new virus strains. We proposed the SEIRD-H model, a modified SEIR model with additional compartments, to simulate the epidemic from incomplete statistical data. We introduce a smoothed stepwise function to model quarantine measures. The proposed approach is verified by cross-validation on historical data and is shown to outperform the baselines. Finally, we extend the obtained model to simulate two strains of COVID-19 and forecast the impact of B.1.1.7 strain on the city of Moscow. Experimental results indicate that B.1.1.7 has the potential to cause a new wave of infections in Moscow that peaks in September-October 2021. 

Narrowing the risk estimate is subject to further study. One approach to improving the forecast is to find the true number of current B.1.1.7 cases via genetic sequencing. The other approach is to model the effect of vaccinations. Currently our model provides a starting point for estimating the effects of new virus strains and can be used to assess risks of other COVID-19 strains and even different epidemics.

The interpretations, conclusions, and recommendations in this work are those of the authors and do not necessarily represent the views of associated organizations.

%
%
\bibliographystyle{IEEEtran}
\bibliography{references} 

\newpage 

\appendices
\section{Two-strain model ODE system}
\label{appendix:ode_systems}
The ODE system for the SEIRD-H one-strain model:
\begin{equation} \label{eq:seridh_one_strain}
    \begin{split}
        \frac{dS}{dt} & = -\frac{\beta S(t) (I(t) + I_v(t))}{N} \\
        \frac{dE}{dt} & = \frac{\beta S(t) (I(t) + I_v(t)) }{N} - \delta E(t) \\
        \frac{dI}{dt} & = (1 - p_i) \delta E(t) - \gamma (1 - \alpha) I(t) - \gamma \alpha I(t)\\
        \frac{dI_v}{dt} & = p_i \delta E(t) - \gamma (1 - \alpha) I_v(t) - \gamma \alpha I_v(t) \\
        \frac{dR}{dt} & = \gamma (1 - \alpha) I(t) \\
        \frac{dR_v}{dt} & = \gamma (1 - \alpha) I_v(t) \\
        \frac{dD}{dt} & = (1 - p_d) \gamma \alpha I(t) \\
        \frac{dD_v}{dt} & = p_d \gamma \alpha I(t) + \gamma \alpha I_v(t) \\
    \end{split}
\end{equation}
The ODE system for the SEIRD-H two-strain model:
\begin{equation} \label{eq:seridh_two_strains}
    \begin{split}
        \frac{dS}{dt} & = -\frac{\beta_1 S(t) (I_1(t) + I_{v_1}(t)) }{N} -\frac{\beta_2 S(t) (I_2(t) + I_{v_2}(t))}{N} \\
        \frac{dE_1}{dt} & = \frac{\beta_1 S(t) (I_1(t) + I_{v_1}(t))}{N} - \delta E_1(t) \\
        \frac{dI_1}{dt} & = (1 - p_i) \delta E_1(t) - \gamma (1 - \alpha) I_1(t) - \gamma \alpha I_1(t)\\
        \frac{dI_{v_1}}{dt} & = p_i \delta E_1(t) - \gamma (1 - \alpha) I_{v_1}(t) - \gamma \alpha I_{v_1}(t) \\
        \frac{dE_2}{dt} & = \frac{\beta_2 S(t) (I_2(t) + I_{v_2}(t))}{N} - \delta E_2(t) \\
        \frac{dI_2}{dt} & = (1 - p_i) \delta E_2(t) - \gamma (1 - \alpha) I_2(t) - \gamma \alpha I_2(t)\\
        \frac{dI_{v_2}}{dt} & = p_i \delta E_2(t) - \gamma (1 - \alpha) I_{v_2}(t) - \gamma \alpha I_{v_2}(t) \\
        \frac{dR}{dt} & = \gamma (1 - \alpha) I_1(t)  + \gamma (1 - \alpha) I_2(t) \\
        \frac{dR_v}{dt} & = \gamma (1 - \alpha) I_{v_1}(t) + \gamma (1 - \alpha) I_{v_2}(t) \\
        \frac{dD}{dt} & = (1 - p_d) \gamma \alpha I_1(t) + (1 - p_d) \gamma \alpha I_2(t) \\
        \frac{dD_v}{dt} & = p_d \gamma \alpha I_1(t) + \gamma \alpha I_{v_1}(t) +  p_d \gamma \alpha I_2(t) + \gamma \alpha I_{v_2}(t) \\
    \end{split}
\end{equation}


\onecolumn
\section{Two-strain SEIRD-H schematic} 
\label{appendix:seird_two_strain}

\begin{figure*}[!h]
    \centering
    \includegraphics[width=\linewidth]{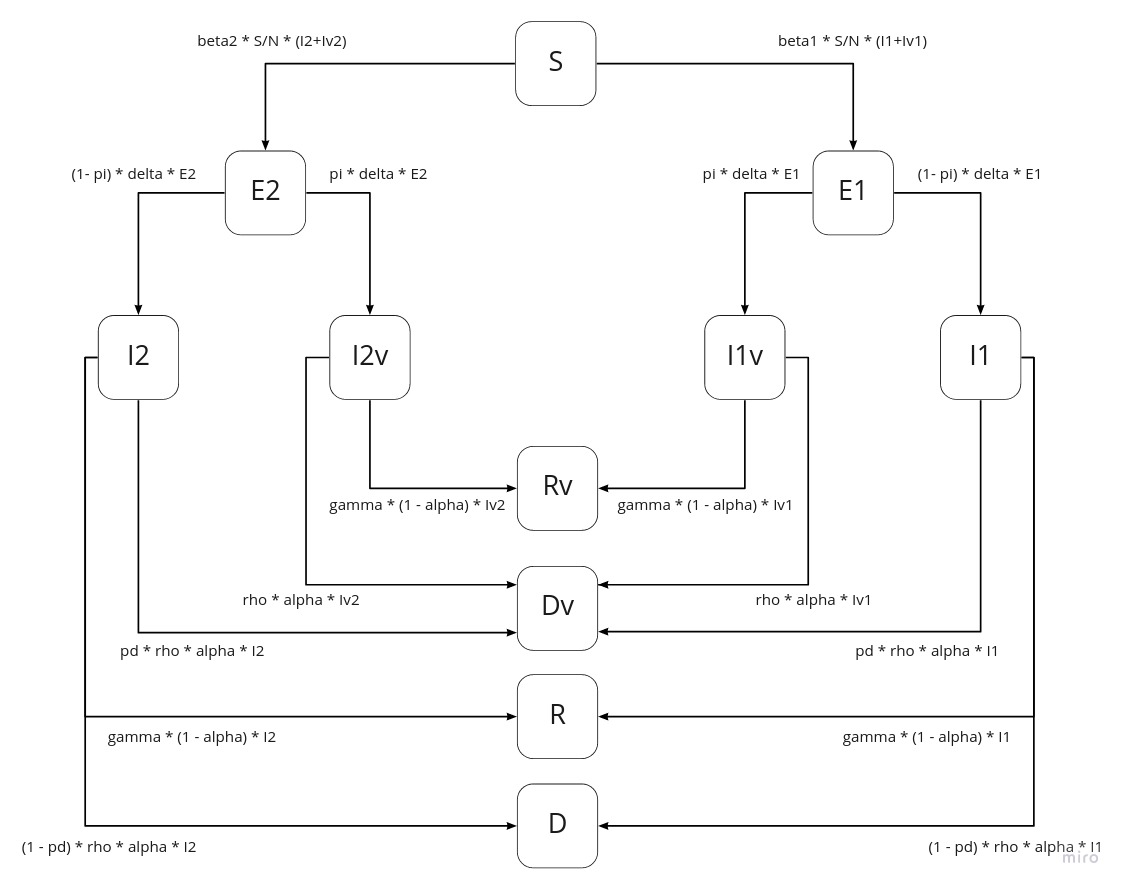}
    \caption{SEIRD-H for two strains.}
    \label{fig:seird_two_strains}
\end{figure*}

\ifCLASSOPTIONcaptionsoff
  \newpage
\fi

\end{document}